\documentstyle[12pt]{article}
\begin{document}
\begin{center}
{\Large\bf Massive States in Chiral Perturbation Theory}\\[1 cm]
\end{center}
\begin{center}
S. Mallik\\
Saha Institute of Nuclear Physics\\
1/AF, Bidhannagar, Calcutta 700064, India
\end{center}
\vspace{1cm}
\begin{abstract}
It is shown that the chiral nonanalytic terms generated by $\Delta_{33}$
resonance in the nucleon self-energy is reproduced
in chiral perturbation theory by perturbing
appropriate local operators contained in the pion-nucleon effective
Lagrangian itself.
\end{abstract}
\newpage
\noindent{\bf 1 Introduction}
\bigskip

The notion of spontaneously broken chiral symmetry [1] as applied to strong
interactions has a long history [2]. Through an important series of
investigations, it has developed into the chiral perturbation
theory ($\chi PT$) based on the
symmetries of  QCD . It has now led to a rather detailed and
quantitative understanding of hadron physics in the low energy region,
not accessible to a conventional perturbative treatment [3].

Consider QCD with u and d quarks only. Its effective Lagrangian
in $\chi PT$ consists of a series of terms in powers of symmetry
breaking parameters (masses) and derivatives (momenta), built out of the
(Goldstone) pion fields incorporating the symmetries. It, however,
brings in a number
of effective coupling constants which, unless estimated from experiment,
 tends to limit the predictive power of this approach [4].
Unitarity is restored up to a certain order
 by including loop diagrams with lower
order terms in ${\cal L}$ as vertices. These loop diagrams give rise to the
nonanalytic contributions, which are proportional to odd powers or
logarithm of the pion mass. They arise from the infrared properties of
Feynman diagrams in the chiral limit.

The baryon fields can also be included in the framework of the effective
field theory [5,6]. Thus the effective Lagrangian for the pion-nucleon system
contains the pion and the nucleon fields. Predictions of $\chi PT$ for
this system can be confronted with rich experimental data.

A question arises as to how the effective Lagrangian takes into account
the contributions (in particular, its nonanalytic pieces)
 of more massive particles whose fields are not present in
it. In a dispersion theoretic framework such particles can be present as
intermediate states. But in the $\chi PT$ approach these
contributions are implicitly contained in the terms of the effective
Lagrangian together with the loops they generate. In other words, the method
not only accounts for higher order interactions of particles whose fields are
present in the effective Lagrangian but also the interactions with the massive
states.

The $\Delta_{33} (1232)$ resonance in the $\pi N$  system is a good example to
discuss the role of higher mass states in the effective field theory. Here we
calculate the contribution of $\Delta_{33}$ exchange
 to the nucleon self-energy loop and extract the
leading nonanalytic piece in it. The
calculation is conveniently carried out in the heavy nucleon formalism.
We then identify the lower order terms in ${\cal L}$
which when treated as vertices in  self-energy loop, reproduce the same
nonanalytic piece.

Of course, the nonanalytic piece is identified in the chiral limit
where the pion mass
$m_{\pi} \rightarrow 0$. But in the real world the difference $\delta$
between the $\Delta$ and $N$ masses is comparable to $m_{\pi}$.
In such cases one may have to sum the contributions from an entire
 series of terms of the effective Lagrangian, which,
in effect, amounts to having nonlocal terms in ${\cal L}$. It may then be
useful to introduce an independent field for the excited state (here
 $\Delta_{33}$ ) to avoid this nonlocality.

In Sec.2 we review the $\chi PT$ in the heavy nucleon formalism. In  Sec.3
we calculate the self-energy diagram in this formalism to extract the
nonanalytic piece.We then identify the local operators in $\cal L$ which
in loops just reproduce this result. Our concluding remarks are
contained in Sec.4.

\vspace{.8cm}

\noindent{\bf 2 Heavy Nucleon $\chi PT$}

\bigskip

Here we use the heavy nucleon formalism of Georgi [7], as applied by
Jenkins and Manohar [8,9] to construct the effective chiral Lagrangian.
 It extends
the usual power counting of $\chi PT$ in presence of nucleon. Here the nucleon
momentum is written as
\begin{equation}
P^{\mu}=mv^{\mu}+p^{\mu}.
\end{equation}
If the nucleon mass $m$ is large, its velocity $ v^{\mu}$ remains (almost)
unchanged by scattering with fixed (small) momentum transfer $p^{\mu}$. One
introduces the nucleon field $b(x)$ with definite velocity $v^{\mu}$
 related to
the original field $B(x)$ by
\begin{equation}
b(x)=e^{imv\!\!\!/ v.x} B(x)
\end{equation}
The reduced field satisfies the modified Dirac equation
\begin{equation}
i\partial\!\!\!/ b(x)=0
\end{equation}
without a mass term. The $\partial ^{\mu}$ produces a $p^{\mu}$ rather
than $P^{\mu}$.
The Dirac structure of the theory simplifies considerably [8]; the bilinear
covariants can now be simply written in terms of $v^{\mu}$ and the spin-vector
$S^{\mu}={i\over 2}\sigma^ {\mu \nu }\gamma_5 v_{\nu}$, which is the
Pauli-Lubanski 4-vector in this formalism.

We briefly review the construction of the effective Lagrangian [8,9].
One introduces the matrix
\begin{equation}
U(x)=e^{i\phi (x)},\   \phi (x)=\tau ^a \phi ^a (x)/{f_{\pi}},\  a=1,2,3,
\end{equation}
where $ \phi ^{a}$  are the pion fields , $f_{\pi}$ the pion decay constant
and $\tau^{a}$ are the Pauli matrices.
Under $ SU(2)_{L}\times SU(2)_{R}, U\rightarrow V_{L} U V_{R}^\dagger,$
where $V_{L}$
and$ V_{R}$ are global SU(2) transformations. The square root of $ U$ is
denoted by $ u,\  U=u^2$. It transforms as
$ u\rightarrow V_{L}u R^{\dagger}=RuV_{R}^{\dagger},  $
defining R implicitly. The reduced nucleon field transforms as
$b(x) \rightarrow   Rb(x).$

With $ u(x)$ one can build two vector fields. One is the gauge field \\
$V_{\mu}={1\over 2}(u\partial_{\mu}u^{\dagger}
 +u^{\dagger} \partial_{\mu} u)$
 and the other is the
axial vector field\\  $A_{\mu}={i\over 2}(u\partial_{\mu}u^{\dagger}
-u^{\dagger} \partial_{\mu}u)$.
The covariant derivative of the reduced nucleon field $b(x)$ is then
$D_{\mu} b=\partial_{\mu} b + V_{\mu} b$.

The lowest order effective Lagrangian is given by
\begin{equation}
{\cal L} ^{(1)}=i\bar{b} v\cdot D b +2g_{\pi N}\bar{b}  S\cdot  A b
\end{equation}
At the next higher order an independent set of operators consists of [10]
\begin{eqnarray}
O_1&=&\bar{b} D^2 b,\nonumber\\
O_2&=&\bar{b} (v \cdot D)^2 b,\nonumber\\
O_3&=&\bar{b} A^2 b ={1\over{16}}\bar{b} b \partial_{\mu}\phi ^{a} \partial
^{\mu} \phi^{a}+\cdots\nonumber\\
O_4&=&\bar{b} (v\cdot A)^2 b={1\over{16}}\bar{b} b (v\cdot \partial \phi^{a})
(v\cdot \partial \phi ^{a})+\cdots \nonumber\\
O_5&=&\bar{b} v\cdot A  s\cdot D b,\nonumber\\
O_6&=&\bar{b} s\cdot D  v\cdot A b,\nonumber\\
O_7&=&\epsilon _{\mu \nu \lambda \sigma} v^{\lambda}\bar{b} S^{\sigma}
A^{\mu} A^{\nu} b \nonumber\\
O_8&=& \bar{b}(uMu+u^{\dagger}Mu^{\dagger})b,
\end{eqnarray}
$M$ being the quark mass matrix,
\[M=\left(\begin{array}{c c}
    m_{u}  &  0  \\
    0      &  m_{d}
    \end{array}\right)\]
In the above enumeration each operator represents a series of terms in
powers of the meson field and the derivative. We have noted above for
later use that only $O_3$ and $O_4$ contain terms which are quadratic
in both these quantities and multiplied by $\bar{b}b$
( not $\bar{b}\tau^{a}b$ ).

\vspace{.8cm}

\noindent{\bf 3 Self-energy loop with $\Delta_{33}$}

\bigskip

We now calculate the contribution of the $\Delta_{33}$ intermediate
state to the nucleon self-energy. From (5) we get the nucleon
propagator as
\begin{equation}
i\over{p\cdot v +i\eta}
\end{equation}
The $\Delta_{33}$ is described by a spinor-vector field $\psi_{\mu}(x)$
. Again the nucleon mass is extracted as in (2) to define a reduced
field $\chi_{\mu} (x)$, which satisfies the free equation of motion,
\begin{equation}
(iv\cdot \partial -\delta)\chi_{\mu}=0,\ \ \delta=m_{\Delta}-m
\end{equation}
The original subsidiary condition $\gamma^{\mu} \psi _{\mu}=0$ eliminating
extra components translates for $\chi_{\mu}$ into
\begin{equation}
v^{\mu} \chi_{\mu}=0, \ \ S^{\mu} \chi_{\mu}=0
\end{equation}
Taking these conditions into account, the $\Delta_{33}$ propagator is
given by
\begin{equation}
iP_{\mu \nu}\over {p\cdot v-\delta +i\eta},
\end{equation}
where
\[P_{\mu \nu}={2\over 3}(v_{\mu} v_{\nu}-g_{\mu \nu}-i\epsilon_{\mu \nu
 \lambda \sigma} v^{\lambda} S^{\sigma})\]
The $\pi N \Delta$ interaction is given by
\begin{equation}
{\cal L_{I}} =ig(\bar{\chi}^{\mu,a} b\partial_{\mu}\phi^{a}-
\bar{b} \chi^{\mu,a} \partial_{\mu} \phi^{a})
\end{equation}

The nucleon self-energy to second order (Fig.1a) at the nucleon pole
$(p\cdot v=0)$ is
\begin{equation}
\Sigma=-3g^2\int {d^4 k\over{(2\pi)^4}}{ i\over{v\cdot k-\delta}}
{i\over{k^2 -m_{\pi}^2}} k^{\mu} k^{\nu} {2\over 3}
(v_{\mu} v_{\nu}-g_{\mu \nu})
\end{equation}
It can be evaluated as in Ref [9].
The $k$ integration in the dimensional regularization scheme results in
two terms. The one proportional to $v_{\mu} v_{\nu}$ goes to zero on
contracting with $(v_{\mu} v_{\nu} -g_{\mu \nu})$, while the other
proportional to $g_{\mu \nu}$ gives
\begin{equation}
\Sigma={{i \pi^{d/2}}\over{(2\pi)^{d}}}\Gamma (1-d/2) 6g^2\int_{0}^{\infty}
{{d\lambda}\over{(\lambda^2+2\lambda \delta +m_{\pi}^2)^{1-d/2}}}
\end{equation}
The integral may be evaluated by integrating by parts to get finally
\begin{eqnarray}
\Sigma &=& {{ig^2}\over{8\pi^2}}\mu^{-2\epsilon}
[(-{1\over{\epsilon}} +\gamma-1-\ln4\pi)\delta (2\delta^2-3m_{\pi}^2)
-{2\over3}\delta (5\delta^2-6m_{\pi}^2)\nonumber\\
&+& \delta (2\delta^2-3m_{\pi}^2)\ln{m_{\pi}^2\over {\mu}^2}
-2(\delta^2-m_{\pi}^2)^{3/2}\ln({{\delta-\sqrt{\delta^2-m_{\pi}^2}}\over
{\delta+\sqrt{\delta^2-m_{\pi}^2}}})]
\end{eqnarray}
where $\mu$ is the renormalization scale, $\epsilon=2-d/2$ and
$\gamma =.5772$.
 The renormalization can can be carried out in the standard way. Here we are
interested in the leading nonanalytic piece in the chiral limit, which
is left untouched by any renormalisation prescription. On expanding the
last term in (14) for $m_{\pi}\ll \delta$, we get the leading nonanalytic
contribution to $\Sigma$ as
\begin{equation}
\Sigma_{nonanalytic}=-{{i3g^2}\over{32\pi^2}} {m_{\pi}^4\over{\delta}}
\ln{m_{\pi}^2}
\end{equation}

Let us now consider the situation as the chiral limit is approached.
Then the relevant internal momenta are of order $m_{\pi}$, so that
$v\cdot p\ll\delta$. We may then expand the $\Delta_{33}$
propagator and retain the leading term in the integrand of (12).
 We thus get the tadpole diagram (Fig.1b) due to insertion of a local operator
generated by the $\Delta_{33}$ intermediate state. The corresponding
self-energy is given by
\begin{eqnarray}
\Sigma _{tadpole} &=& -{3g^2\over {\delta}}\int {d^4k\over{2\pi ^4}}
{1\over{k^2-m_{\pi}^2}}
k^{\mu} k^{\nu}({2\over3}v_{\mu} v_{\nu}-g_{\mu \nu})\\
& = & {i3g^2\over{32\pi^2}}{m_{\pi}^4\over \delta}\mu ^{-2\epsilon}
({1\over\epsilon}-\gamma+{3\over2}+\ln{4\pi}-\ln{m_{\pi}^2\over\mu ^2})
\end{eqnarray}
which coincides with (15) for the chiral logarithm obtained from the
Feynman diagram with $\Delta_{33}$ intermediate state. Further, looking at
the list (6) of $O(k^2)$ operators, the one in (16) is easily identified
as
\[{32g^2\over\delta}(O_4- O_3)\]

\vspace{.8cm}

\noindent{\bf 4 Conclusion}

\bigskip

We demonstrate, within the context of the nucleon self-energy,
 that nonanalytic terms in the chiral limit produced by loops
involving higher mass states can be reproduced exactly by perturbing
appropriate local operators contained in the effective chiral
Lagrangian itself. This result should be true in general, as the
manipulations involved here are of the same kind as required in
 establishing the short distance expansion [11]. It follows that the
chiral properties of $\pi N$ system can be adequately described by an
effective chiral Lagrangian involving pion and nucleon fields alone.

Our work disproves conclusively a claim made in a recent work [12] that
the leading nonanalytic piece in the chiral limit generated by $\Delta_{33}$
exchange in $\pi N$ $\sigma$-term [13] cannot be reproduced by a
phenomenological Lagrangian for this system. What these authors fail to
note that all the relevant pieces of this Lagrangian must be included in
the loop calculation; otherwise unitarity will be violated.

There is, however, a question of quantitative importance of the nonanalytic
pieces in the real world. As mentioned in the introduction, if an excited
 state is low enough, the $\chi PT$ involving the basic fields only
 may not be very convenient. Indeed this has been argued to be the case
for $\Delta_{33}$ itself [9]. In that case, one may include the
field corresponding to such a state in constructing the effective
Lagrangian. But to be consistent one must calculate all quantities within
the framework of $\chi PT$ , as was pointed out long ago in Ref.[14,15].

I am very thankful to H. Leutwyler for help and comments.
I also wish to thank J. Gasser and U. -G. Meissner for discussions. I
acknowledge the kind hospitality at the University of Berne, where the
work had started.

\newpage

\noindent{\bf Figure Caption}

\bigskip

Fig.1. Nucleon self-energy diagrams. The thin and thick solid lines and
the dashed line represent the nucleon, the $\Delta_{33}$ resonance and
the pion respectively. (a)  $\Delta_{33}$ exchange. (b) insertion of
local operators (represented by the dot) stated in the text.

\begin{thebibliography}{99}
\bibitem{10} We have used equation of motion for b obtained by varying
${\cal L}^{(1)}$ to reduce the number of independent $O_{i}$ s in (5).
\end{thebibliography}
\end{document}